\documentclass[aps,pre,reprint,groupedaddress,fleqn]{revtex4-2}
\usepackage[utf8]{inputenc}

\usepackage{amsmath}
\usepackage{amssymb}
\usepackage{amsfonts}
\usepackage{graphicx}
\usepackage[unicode=true,
 bookmarks=true,bookmarksnumbered=false,bookmarksopen=false,
 breaklinks=false,pdfborder={0 0 1},backref=false,colorlinks=true]
 {hyperref}
\hypersetup{
 linkcolor=magenta, urlcolor=blue, citecolor=blue, pdfstartview={FitH}, unicode=true}
\usepackage{xcolor}

\begin{document}

\title{Self-averaging of digital memcomputing machines}
\author{Daniel Primosch}
\affiliation{Department of Physics, University of California San Diego, La Jolla, CA 92093}

\author{Yuan-Hang Zhang}
\affiliation{Department of Physics, University of California San Diego, La Jolla, CA 92093}

\author{Massimiliano Di Ventra}
\email{email: diventra@physics.ucsd.edu}
\affiliation{Department of Physics, University of California San Diego, La Jolla, CA 92093}

\begin{abstract}
   
Digital memcomputing machines (DMMs) are a new class of computing machines that employ non-quantum dynamical systems with memory to solve combinatorial optimization problems. Here, we show that the time to solution (TTS) of DMMs follows an inverse Gaussian distribution, with the TTS {\it self-averaging} with increasing problem size, irrespective of the problem they solve. We provide both an analytical understanding of this phenomenon and numerical evidence by solving instances of the 3-SAT (satisfiability) problem. The self-averaging property of DMMs with problem size implies that they are increasingly insensitive to the detailed features of the instances they solve. This is in sharp contrast to traditional algorithms applied to the same problems, illustrating another advantage of this physics-based approach to computation.  

\end{abstract}
\maketitle

\section{Introduction}
Combinatorial optimization problems emerge in a wide variety of applications in both academia and industry~\cite{complexity-bible}. They require finding an assignment of variables---which typically take the values of logical 1 or logical 0---that satisfies a set of constraints. Given one of these problems, 
e.g., the satisfiability (SAT) with exactly three variables per constraint (known as 3-SAT), one can use different algorithms to find the solution 
of its instances. Since the various instances of a problem are different, even at a fixed number of variables and constraints, and one can choose different initial conditions for the algorithm, one expects the time to solution (TTS)---namely the 
time it takes an algorithm to solve those instances---to vary according to some distribution. 

If the problem to solve is hard, such as the 3-SAT, both the mean $\mu$ and the standard deviation $\sigma$ of TTS are expected to increase exponentially with increasing problem size, $N$. We then expect $\sigma^2(N)/\mu^2(N)\sim O(1)$ in the limit $N\rightarrow \infty$.
We would then say that in this case the TTS is {\it not self-averaging}: its relative variance does not go to zero 
in the thermodynamic limit~\cite{vanKampen}. (We will provide an explicit example of this phenomenon for traditional algorithms below.)

In recent years a new computing paradigm, dubbed memcomputing~\cite{diventra13a,UMM,MemComputingbook}, has been suggested which employs {\it time non-locality} (memory) to tackle computational problems. In particular, its digital version (digital memcomputing machines or DMMs)~\cite{DMM2}, 
has been designed to solve problems in the combinatorial optimization class. Unlike traditional algorithms, DMMs map the original problem into nonlinear (nonquantum) dynamical systems whose point attractors are the solutions (if they exist) of such a problem. Since DMMs do not support chaos~\cite{no-chaosa} (as well as quasiperiodic orbits), their ordinary 
differential equations (ODEs) can be efficiently simulated and tested on our traditional computers, even before their hardware implementation. These simulations have already shown substantial advantages over state-of-the-art algorithms on a wide range of applications (see, e.g.,~\cite{MemComputingbook} for a sample of problems already tackled with this paradigm). 

In this paper, we show that the physics of DMMs implies that their TTS follows an inverse Gaussian distribution~\cite{heur_deriv_inv_gauss_1987}, irrespective of the problem they solve. Most importantly, the TTS {\it self-averages} with increasing problem size, namely $\sigma^2(N)/\mu^2(N)\rightarrow 0$ as  $N^{-\theta}$, for $N\rightarrow \infty$, with $\theta =1$ (``strong self-averaging'') for physical noise and $\theta < 1$ (``weak self-averaging'') for numerical noise. Apart from an analytical understanding of this phenomenon, we corroborate this prediction by numerically solving satisfiable instances of the 3-SAT~\cite{barthel2002} in the presence of physical noise, and contrast it to the TTS of several well-known traditional algorithms~\cite{computational-complexity-book, sorensson2005minisat, fleury2020cadical}.
The self-averaging property of DMMs with problem size implies that they are increasingly {\it insensitive} to the detailed features of the instances they solve. These results both clarify the physics behind these machines and illustrate their advantage in computing. 

\section{DMMs for 3-SAT}
Without loss of generality we will focus on the 3-SAT (with instances taken from~\cite{barthel2002}), which is a collection of clauses (OR gates) with exactly three Boolean variables, $y_i$ ($i=1,2,3$), with the clauses related by logical conjunctions (AND gates). A DMM (not necessarily unique) for such a problem can be constructed as follows~\cite{Sean3SAT}. The variables $y_i$ are first transformed into continuous ones, $v_i\in [-1, 1]$ (which could be voltages in an actual circuit realization of DMMs~\cite{DMM2}), where $v_i>0$ corresponds to $y_i=1$, and $v_i<0$ corresponds to $y_i=0$. The $m$-th clause is represented by $(l_{m, i}\!\vee l_{m, j}\!\vee l_{m, k})$, where $l_{m, i}=\bar{y}_i$ or $y_i$  depending on whether or not $y_i$ is negated, and $\vee$ is the symbol of the OR gate. Each Boolean clause is then mapped into a continuous constraint function,
\begin{equation}
	C_m(v_i, v_j, v_k) = \frac{1}{2}\min\left(1-q_{m, i}v_i, 1-q_{m, j}v_j, 1-q_{m, k}v_k\right),
\end{equation}
where $q_{m, i}=1$ if $l_{m_i}=y_i$ and $q_{m, i}=-1$ if $l_{m_i}=\bar{y}_i$. It is clear that the $m$-th clause evaluates to true if and only if $C_m<1/2$. 

Time non-locality is introduced in the form of additional ``memory degrees of freedom'' which guarantee that the only fixed points of the dynamics correspond to the solution of the problem, and no other critical points, 
other than saddle points, are present in the phase space~\cite{MemComputingbook}. Following~\cite{Sean3SAT}, we introduce two additional memory variables for each OR gate: a ``short-term'' memory $x_{s, m}$ and a ``long-term'' memory $x_{l, m}$. The dynamics of DMMs for a 3-SAT instance with $N$ variables and $M$ clauses are then ($i=1,\dots, N$; $m=1,\dots, M$): 
\begin{equation}
			\dot{v}_{i}=\sum_{m} x_{l, m} x_{s, m} G_{m, i}+\left(1+\zeta x_{l, m}\right)\left(1-x_{s, m}\right) R_{m, i},\label{eq:DMM}
\end{equation}
\begin{flalign}
	\begin{split}
		&\dot{x}_{s, m}=\beta\left(x_{s, m}+\epsilon\right)\left(C_{m}-\gamma\right),\quad {x}_{s, m} \in[0,1],\\
		&\dot{x}_{l, m}=\alpha\left(C_{m}-\delta\right),\quad {x}_{l, m} \in[1,10^4M],
	\end{split}\label{eq:DMMx}
\end{flalign}
 where the ``gradient-like'' term $G_{m, i}=\frac{1}{2}q_{m, i}\min(1-q_{m, j}v_j, 1-q_{m, k}v_k)$, while the ``rigidity'' term $R_{m, i}=\frac{1}{2}(q_{m, i}-v_i)$ if $C_m=\frac{1}{2}(1-q_{m, i}v_i)$, and $R_{m, i}=0$ otherwise. As in~\cite{Sean3SAT} we have chosen the parameters $\alpha=5, \beta=20, \gamma=0.25, \delta=0.05,\epsilon=10^{-3}, \zeta=0.1$.

We refer to~\cite{Sean3SAT,MemComputingbook} for a thorough discussion of how Eqs.~(\ref{eq:DMM}) and~(\ref{eq:DMMx}) have been obtained. Here, we take them at face value as representations of DMMs and point out that they can be compactly written as  $\dot{\bf{x}}(t)=F({\bf x}(t))$, with $\bf{x}$ the collection of continuous variables, $v_i$, and memory variables, $x_{s, m}, x_{l, m}$, and   
$F$ the flow vector field that can be read from the right-hand side of Eqs.~(\ref{eq:DMM}) and~(\ref{eq:DMMx}).  

To make direct contact with experiments, as done in~\cite{Bearden}, we add Gaussian white noise, $l(t)$, to the memory variables of Eqs.~(\ref{eq:DMMx}). $l(t)$ satisfies $\langle l(t)\rangle=0$ and $\langle l(t) l(t')\rangle =\Gamma\delta(t-t')$, where $\Gamma$ is the noise strength. Of course, the errors introduced by the numerical integration of these ODEs also play the role of ``noise'', and we refer to the Supplemental Material (SM) \cite{SM} for the results in the absence of additive noise. 
\begin{figure}[t]
	\centering
	\includegraphics[width = 0.47\textwidth]{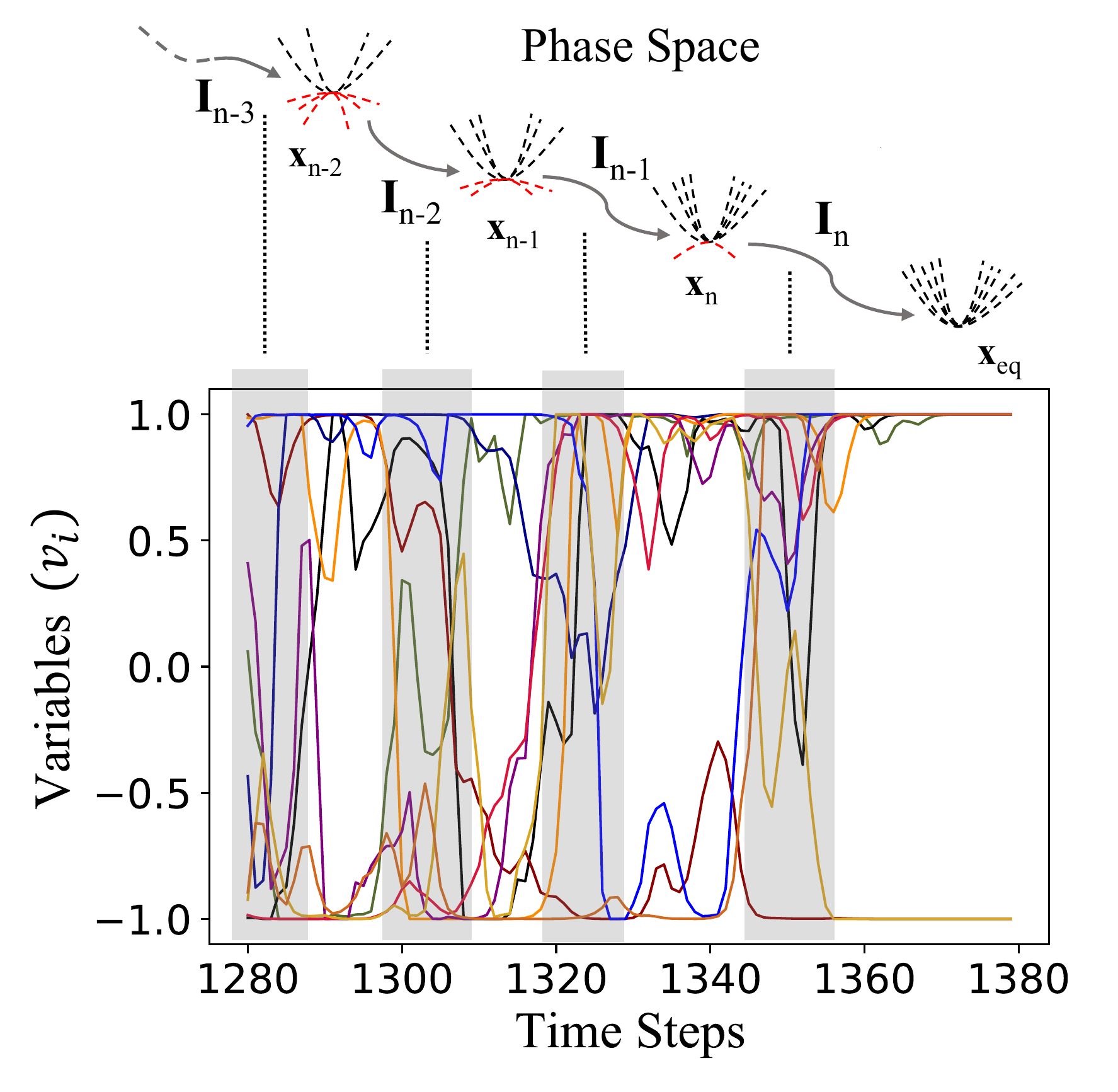}
	\caption{Top panel: Schematic of the state dynamics of a DMM which transitions from a critical point (e.g., ${\bf x}_{n-1}$) to a more stable critical point (${\bf x}_{n}$) in the phase space in such a way as to reduce the number of unstable directions after each instantonic jump (${\bf I}_{n-1}$). Bottom panel: Numerical example of the DMM represented by Eqs.~(\ref{eq:DMM}) and~(\ref{eq:DMMx}) solving a 3-SAT instance with 3,000 variables and 21,000 constraints. The instantonic jumps are the shaded regions where negative (logical 0) variable values suddenly cross to positive (logical 1) ones and vice versa. The area in between corresponds to a critical point.}
	\label{fig:phase space}
\end{figure}

\section{The dynamics of DMMs}
It was shown both analytically (using supersymmetric topological field theory) and numerically that DMMs, with or without additive noise, find a solution to the problem instance by following specific trajectories (``instantons'') in the phase space~\cite{topo,DMtopo}. In particular, DMMs ``transition'' from a critical point (a saddle point for which $F({\bf x}(t))=0$ in Eqs.~(\ref{eq:DMM}) and~(\ref{eq:DMMx})) with a certain number of unstable directions to a more stable critical point, until a solution is found. An example of such dynamics for a 3-SAT instance with 3,000 variables and 21,000 constraints is provided in Fig.~\ref{fig:phase space}, where the sudden jumps of the continuous variables $v_i$ correspond to 
the instantons connecting critical points. These results have been obtained by integrating the DMMs' Eqs.~(\ref{eq:DMM}) and~(\ref{eq:DMMx}) using the forward Euler method~\cite{Sauer} with fixed integration step ($\Delta t=$ 0.2 in arbitrary time units throughout the main text), and no 
additional noise. 
\begin{figure}[t]
	\centering
	\includegraphics[width = 0.49\textwidth]{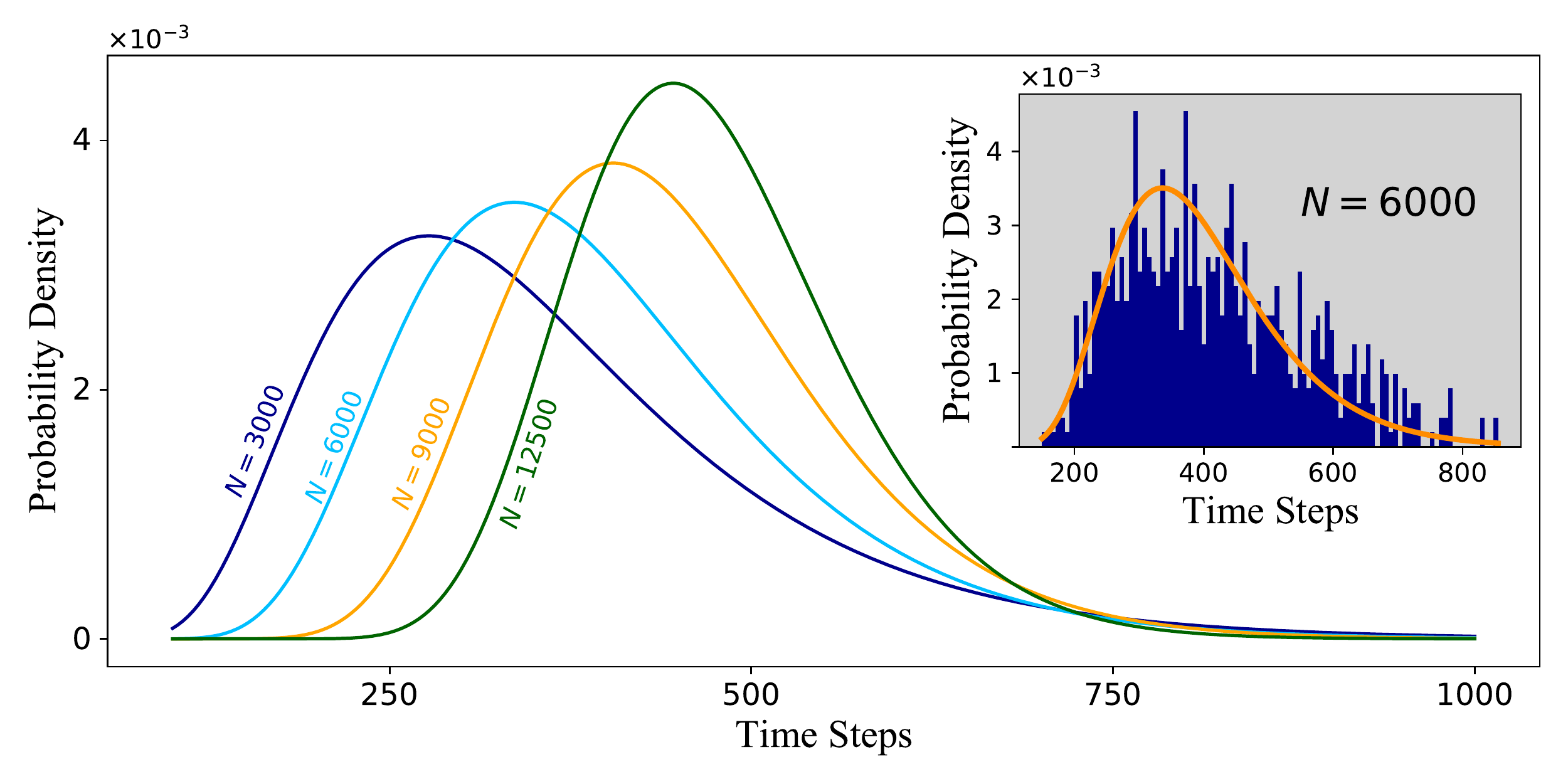}
	\caption{Various fits (inverse Gaussian distributions) of the TTS (in number of integration steps) of DMMs out of 1,000 3-SAT instances (taken from~\cite{barthel2002}) of varying number of variables, $N$, at a fixed clause-to-variable ratio of 7. The noise strength is $\Gamma= 0.12$ (in inverse time units). Inset: the histogram of TTS (in number of integration steps) of the DMM with  $N =$ 6,000 variables and 42,000 clauses. The fit is an inverse Gaussian distribution, Eq.~(\ref{inverseG}), with 
	$\lambda = 3720\pm420$ and $\mu = 390\pm10$.}
	\label{fig:example_fit}
\end{figure}

\subsection{Distribution of TTS for DMMs}
As anticipated, in the presence of noise (whether physical and/or numerical) we expect the TTS to follow some distribution. In order to determine what type of distribution we expect---irrespective of the problem to solve---we can draw on a physical analogy from the dynamics of DMMs as depicted in Fig.~\ref{fig:phase space}. The state vector ${\bf x}(t)$ in phase space can be interpreted as ``position of a particle'' which is {\it directed} towards the solution and subject to some noise. The particle then has a ``drift velocity'', $\delta>0$, which is analogous to how strongly the DMM is driven towards the solution of the problem instance. In this physical system, the noise, modeled by a diffusion constant $\nu$, is analogous to the noise a DMM experiences. 

We can then ask how long it takes such a particle to reach an arbitrary distance $L$ (the solution of the problem) from its starting point (the initial condition). We may call this the time to barrier (TTB), which is the physical analog of the TTS of DMMs. This TTB is well known in the literature and is given by an inverse Gaussian probability density function of the form~\cite{heur_deriv_inv_gauss_1987}  
\begin{equation}
 P_{\delta, L, \nu}(t) = \sqrt{\frac{L^2}{2\pi \nu t^3}}e^{-\delta^2(t-\frac{L}{\delta})^2/2\nu t},\label{invG}
\end{equation}
where $t>0$ corresponds to the time for the particle to reach the distance $L$. We can re-write Eq.~(\ref{invG}) with just two independent parameters by making the substitution $
\mu = L/\delta, \lambda = L^2/\nu$. This gives the simplified equation
\begin{equation}
 P_{\lambda, \mu}(t) = \sqrt{\frac{\lambda}{2\pi t^3}}e^{-\lambda(t-\mu)^2/2\mu^2 t}\label{inverseG}.
\end{equation} 
The mean of this distribution is $\mu$, while its variance is $\sigma^2=\mu^3/\lambda$~\cite{heur_deriv_inv_gauss_1987}. If the analogy we have just made holds, we should find a similar distribution for the TTS of DMMs.

We verify this analytical prediction by numerically integrating Eqs.~(\ref{eq:DMM}) and~(\ref{eq:DMMx}), using again the forward Euler method, where we added Gaussian white noise to the memory variables in Eqs.~(\ref{eq:DMMx}) with a strength $\Gamma = 0.12$ (in inverse time units) \footnote{Following~\cite{Bearden}, we can relate the noise strength $\Gamma$ (in s$^{-1}$) to an effective external temperature (in K). The value considered in the present work, $\Gamma = 0.12$ s$^{-1}$, would then correspond to a temperature comparable to room temperature $\sim 300$K.}. The TTS can be simply defined as the amount of integration steps it took the DMM to find a solution. In the inset of Fig.~\ref{fig:example_fit} we report the histogram of TTS for a DMM solving 1,000 3-SAT instances~\cite{barthel2002} with $N = $ 6,000 variables and 42,000 clauses. The histogram is well fitted by the inverse Gaussian distribution~(\ref{inverseG}) with parameters $\lambda = 3720\pm420$ and $\mu = 390\pm10$~\footnote{Note that the TTS distribution~(\ref{inverseG}) arises also for a given instance (namely at {\it fixed} number of variables and clauses) over an {\it ensemble} of initial conditions. See Fig.~S9 in the SM \cite{SM}.}. The 
main panel of Fig.~\ref{fig:example_fit} shows the fits for different numbers of variables with constant clause-to-variable ratio. It indicates a sharpening of the inverse Gaussian distribution with increasing number of variables. (See the SM \cite{SM} for more statistics using different integration methods and for different clause-to-variable ratios.)

\subsection{Self-averaging of TTS}
To understand this better we need to determine how this distribution varies with problem size (the number of variables $N$, if we fix the clause-to-variable ratio). In particular, we want to know how the ratio $\sigma^2/\mu^2$ varies 
with increasing $N$. To do this let us again use the physical analogy we have just developed. 

We can easily relate the distance $L$ from the initial point of the dynamics to the number of variables $N$, by noting 
that $L$ would correspond to the average radius of an $N-$sphere, $S^N(L)$, in phase space. In the limit of large $N$, this is simply $L\sim \sqrt{N}$ \cite{regular_polytopes}. 

The diffusion constant $\nu$ is related to the amount of noise in the system. For {\it physical} noise (such as the additive one we consider here) this is only determined by some external parameter, such as the temperature~\cite{vanKampen}. In other words, for physical noise $\nu$ is {\it independent} of $N$. 

Finally, the dependence of the drift velocity $\delta$ on $N$ can be obtained as follows. If we have $N$ continuous variables $v_i$, the initial energy injected into the system is $E=N \times E_0$, where $E_0$ is the energy associated with a single variable (and at the initial time $t=0$, the variables are independent of each other; they are coupled immediately by the memory variables for $t>0$). Assuming the $N$ variables have all unit mass (the mass value is irrelevant as long as it is a constant), the $N$ dependence of the drift velocity is then $\delta \sim \sqrt{E}=\sqrt{N\times E_0}$. This means that the variance $\sigma^2(N)= (L \nu)/\delta^3\sim \nu/N$. 
\begin{figure}
	\includegraphics[width = 0.47\textwidth]{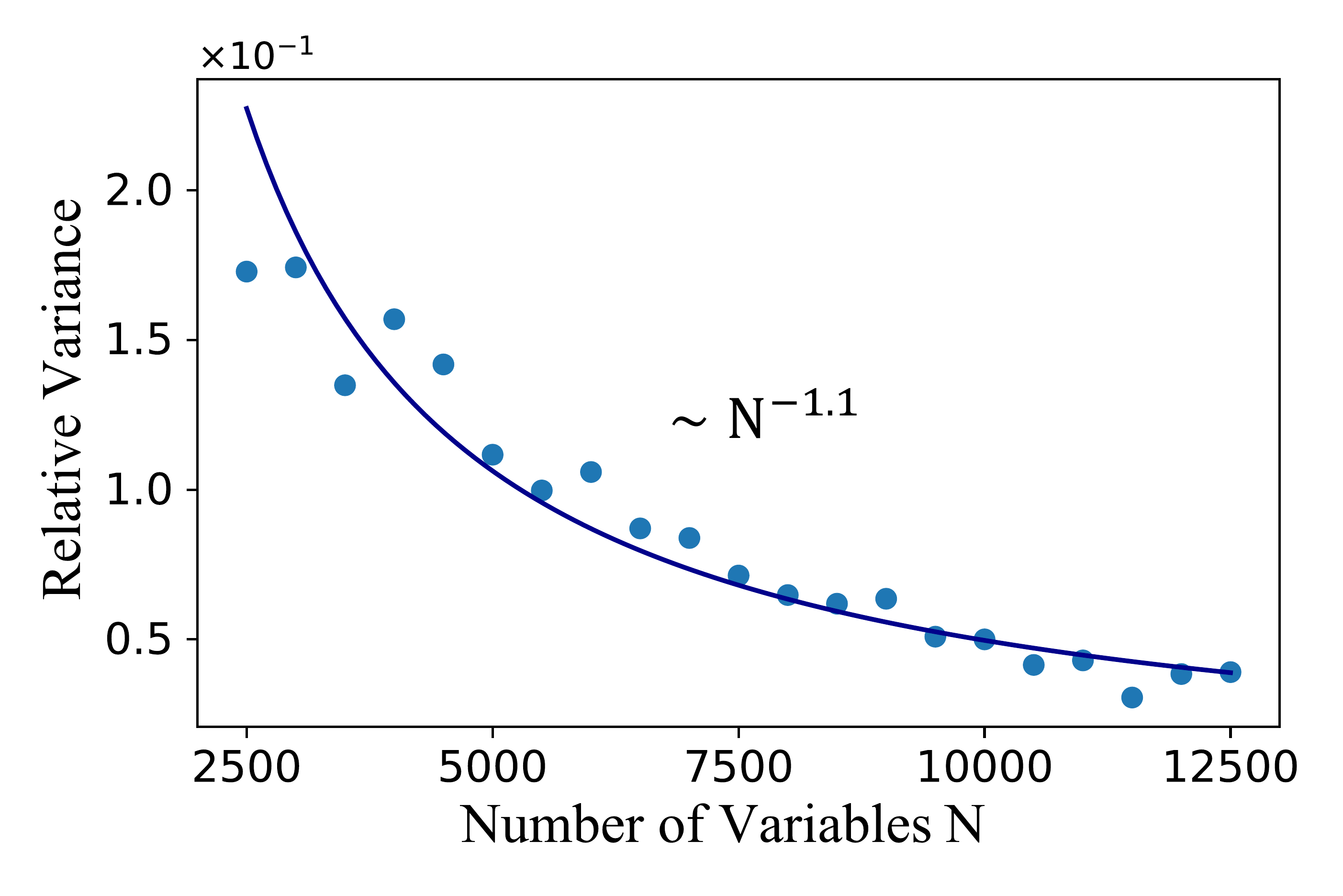}
	\caption{Relative variance $\sigma^2/\mu^2$ as a function of number of variables $N$ of the TTS distributions of the inset of Fig.~\ref{fig:example_fit}. The relative variance is well fitted by a power law $\sigma^2/\mu^2 \sim N^{-1.10\pm0.07}$ consistent with self-averaging behavior.}
	\label{fig:variance}
\end{figure}

By putting all this together, we conclude that the TTS of DMMs subject to physical noise (strongly) {\it self-averages}: $\sigma^2(N)/\mu^2(N)\rightarrow 0$ as $N^{-1}$, for $N\rightarrow \infty$. This is the same result one would obtain for a self-averaging observable in the central limit theorem~\cite{aharony_harris_1996}. It shows that DMMs become increasingly insensitive to the microscopic details of the instances they solve (how the different variables 
are distributed in the various clauses). Note also that the result obtained by this physical analogy is valid {\it irrespective} of the problem the DMMs solve. It is only related to their general dynamical behavior. 

In Fig.~\ref{fig:variance} we show numerical results corroborating the previous analysis, where we solve Eqs.~(\ref{eq:DMM}) and~(\ref{eq:DMMx}) (the latter ones with added noise of strength $\Gamma = 0.12$) for various 
numbers of variables, $N$, and at a fixed clause-to-variable ratio. The relative variance is well fitted by the curve $\sigma^2(N)/\mu^2(N) \sim N^{-1.10\pm0.07}$, showing also that, in this case, the numerical noise does not much affect the 
stochastic dynamics. 

If we had only numerical noise, or the latter strongly couples to the additive noise, the assumption that the diffusion 
coefficient $\nu$ is independent of $N$ would no longer be valid because of the accumulation of errors. In this case, we expect $\nu(N)$ to increase with $N$, with the form of this function dependent on the numerical method that is employed. This is shown in Figs. S7 and S8 of the SM \cite{SM}, where we still find the TTS to be self-averaging but weakly [$\sigma^2(N)/\mu^2(N) \sim N^{0.39\pm0.02}$ for the forward Euler method and $N^{0.29\pm0.02}$ for Runge-Kutta fourth order]. 

\subsection{Non-self-averaging of traditional algorithms}
\begin{figure}[t]
\includegraphics[width=0.47\textwidth]{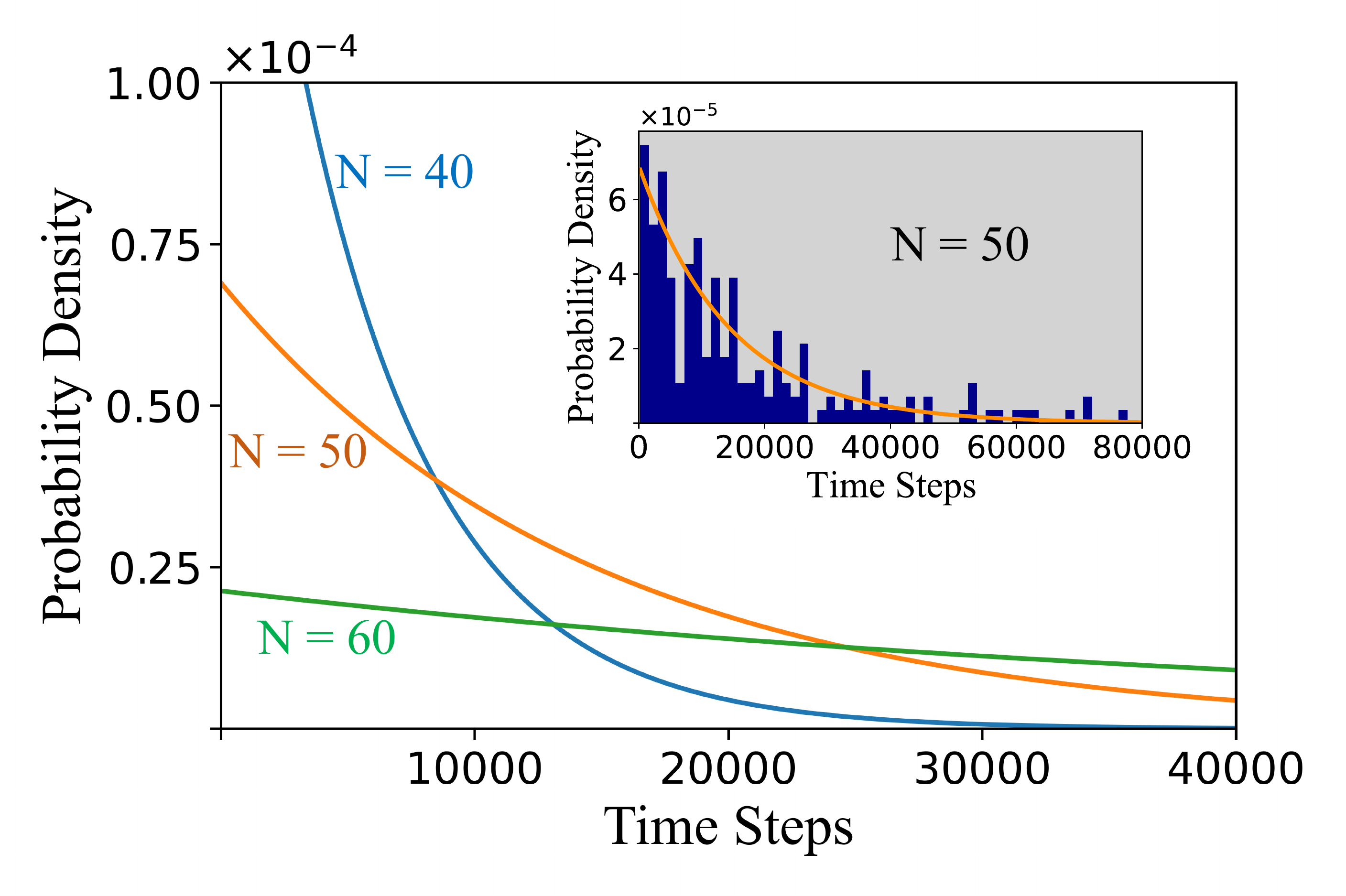}
\caption{The distributions of TTS for the local search algorithm WalkSAT plotted for $N=40, 50, 60$ variables (over 200 instances) at a fixed clause-to-variable ratio of 7. The TTS is well fitted by an exponential probability distribution for all cases considered. Inset: an example for $N=50$ variables of the histogram of TTS used to obtain the fits in the main panel.}
\label{fig:walksat}
\end{figure}

We finally conclude by showing that the self-averaging property of the TTS of DMMs is not shared by traditional algorithms applied to the same problems. We show this explicitly in 
Fig.~\ref{fig:walksat} for a widely used (and representative) local search algorithm (WalkSAT)~\cite{computational-complexity-book} applied to the same instances used previously~\cite{barthel2002}. (In the SM \cite{SM}, we show additional numerical results using two different conflict-driven clause learning algorithms, MiniSAT \cite{sorensson2005minisat} and Kissat \cite{fleury2020cadical}, and we find, as expected, that they are also not self-averaging.) 

This algorithm simply flips one or a few variables at each iteration according to some prescribed rule. Unlike a DMM, the probability distribution of the TTS of this local search algorithm follows well an {\it exponential} fit of the type $P_{\lambda}(t)=\lambda e^{-\lambda t}$. (See the SM \cite{SM} for the analytical justification of why such a 
distribution has to be an exponential.) 
Here, $\lambda$ is the inverse of the average ($\mu=1/\lambda$), and the variance of this distribution is $\sigma^2=1/\lambda^2$. We then see that the relative variance is $\sigma^2/\mu^2 =1$, {\it irrespective} of the size 
of the problem. As anticipated, this traditional algorithm is {\it not} self-averaging.

\section{Conclusions}
We have shown that the TTS of DMMs follows an inverse Gaussian distribution with a relative variance that goes to zero as the size of the problem increases. This is true both in the presence of physical noise 
(showcasing strong self-averaging) as well as numerical noise (weak self-averaging). This means that DMMs are increasingly insensitive to the microscopic details of the instances they solve with increasing problem size. This property is not shared by traditional algorithms, illustrating a substantial advantage of DMMs in the solution of combinatorial optimization problems. 

\section*{Acknowledgments}
This work is supported by NSF under grant No. 2229880. A demo code to reproduce the results in this work can be found 
at \cite{github_link}.

\bibliography{SUSYref}

\clearpage
\newpage
\appendix

\begin{center} 
{\large \bf Supplemental Material: Self-averaging of digital memcomputing machines}
\end{center} 

\setcounter{page}{1}
\setcounter{figure}{0}
\setcounter{equation}{0} 
\makeatletter
\renewcommand{\thefigure}{S\arabic{figure}}
\renewcommand \theequation{S\@arabic\c@equation}
\renewcommand \thetable{S\@arabic\c@table}
\makeatother

\section*{Exponential distribution of local search algorithms}
The time-to-solution (TTS) distribution of a local search algorithm---such as the WalkSAT used in the main text---is known to follow an exponential distribution \cite{SLS_RTD_Hoos_2002}. Here, we present a heuristic derivation of why that is the case.

Since the flip of a particular variable in the problem specification is independent of previous flips and of the other variables, each decision to flip a particular variable is an event described by a discrete Poisson distribution:

\begin{equation}
    P_{Poisson}(k) = \frac{e^{-\lambda}\lambda^k}{k!},
\end{equation}
where $k$ represents the number of rare events, namely those such that a single variable flip results in a solution of the given problem, and $\lambda$ represents the rare event rate. The probability of a variable flip resulting in {\it no solution} is then given by $P(k=0) = e^{-\lambda}$. The fact that variable flips are independent of one another allows us to compute the likelihood that a variable flip resulted in a solution at some time $T$ greater than $t$:
\begin{equation}
    P(T>t) = \prod_{i=1}^t P(k=0)
    = e^{-\lambda t}.
\end{equation}
We then obtain the cumulative distribution function ($f_{CDF}$) via the complement, 

\begin{equation}
    P(T\leq t) = 1 - e^{-\lambda t}
    = f_{CDF}(t).
\end{equation}
The probability density function ($f_{PDF}$) is simply the derivative of $f_{CDF}$: 

\begin{equation}
    f_{PDF}(t) =  \lambda e^{-\lambda t},
\end{equation}

which is the TTS distribution we expected.

\section*{Results using conflict-driven clause learning algorithms}
Many modern SAT solvers are built upon the conflict-driven clause learning (CDCL) algorithm \cite{marques2003grasp}. Unlike WalkSAT, a CDCL solver tries to find a satisfying configuration by assigning truth values to selected variables, propagating the truth values using the Boolean constraints, and backtracking when any conflict is found. 

CDCL solvers typically use many heuristics, and the performance of the solver would depend on the the distribution of the clauses, making it difficult to analytically understand their TTS distribution. Instead, in this section, we only report the numerical results using two popular CDCL solvers: MiniSAT \cite{sorensson2005minisat}, winner of all industrial categories in the SAT 2005 competition and one of the most referenced SAT solvers; and Kissat \cite{fleury2020cadical}, winner of the main tracks of SAT competition 2020 and 2021. As in the main text, we use Barthel instances \cite{barthel2002} at clause-to-variable ratio 7 as test problems. 

\begin{figure}[htbp]
    \includegraphics[width=0.47\textwidth]{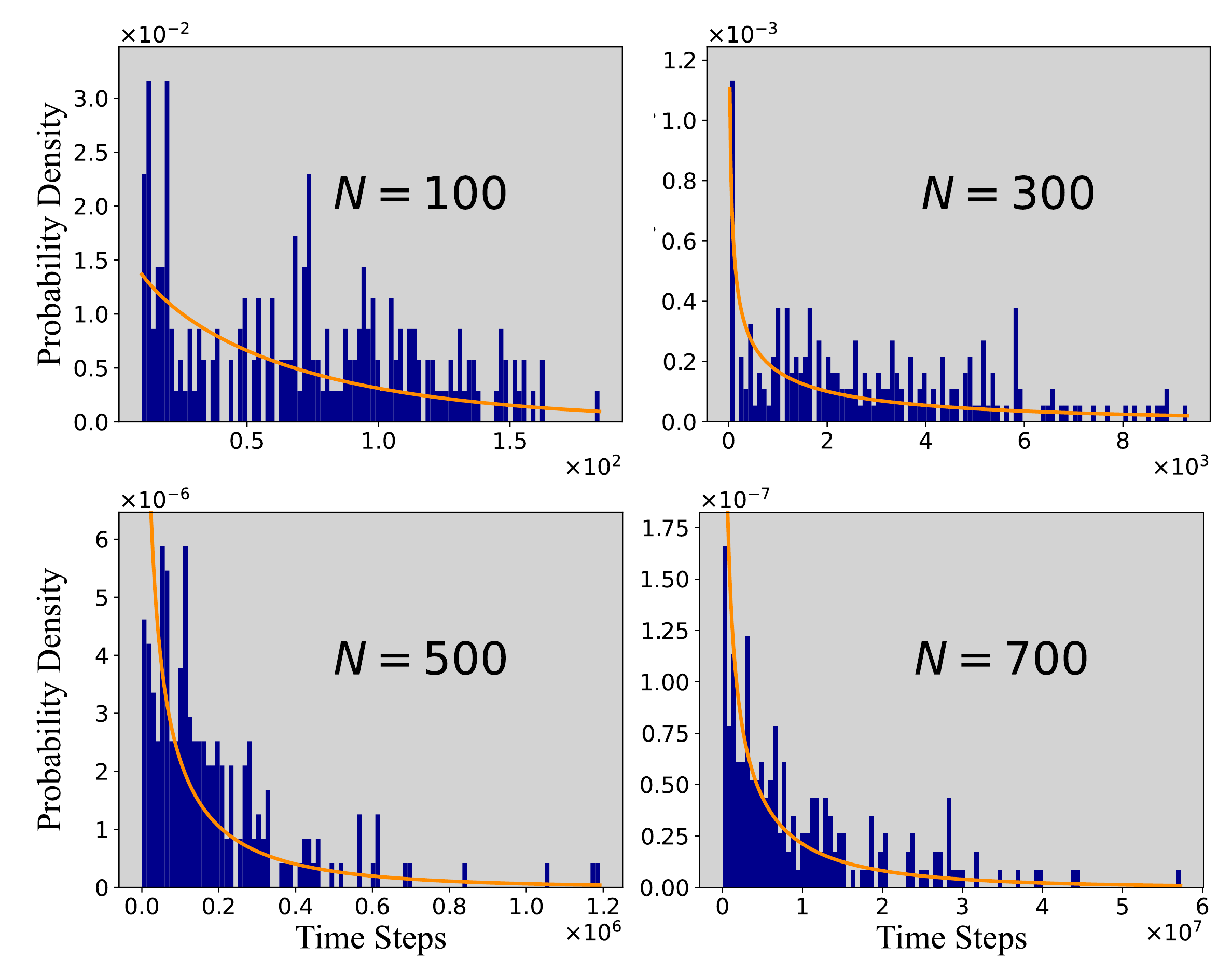}
    \caption{TTS distribution of the MiniSAT algorithm. For each histogram, we attempted the solution of 200 Barthel instances at clause-to-variable ratio 7 and different number of variables $N$. All instances were solved. The histograms can be approximately fitted using a Weibull distribution, Eq.~\eqref{eq:weibull}, with $k<1$. Note how the scales of both axes vastly change with problem size $N$. }
    \label{fig:minisat}
\end{figure}

\begin{figure}[htbp]
    \includegraphics[width=0.47\textwidth]{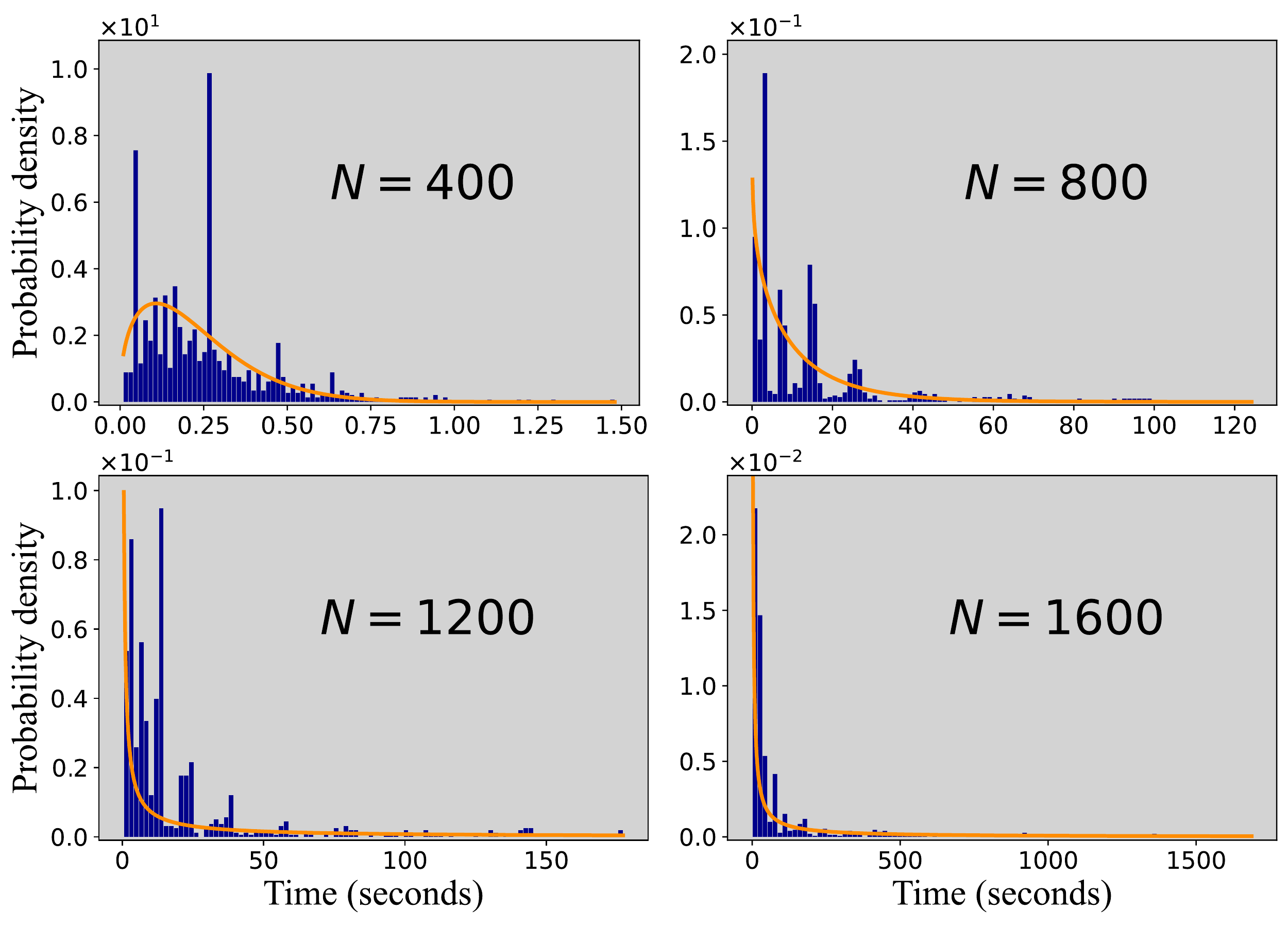}
    \caption{TTS distribution of the Kissat algorithm. For each histogram, we attempted on 999 Barthel instances at clause-to-variable ratio 7, for different number of variables $N$. For $N=400$ and 800, all instances are solved. 995 are solved for $N=1200$, and 997 are solved for $N=1600$. The histograms can be approximately fitted using a Weibull distribution.
    For a clearer visualization, for $N=800, 1200$ and $1600$, only the first $90\%$ of the solved instances are plotted, since the hardest instances solved takes more than $10^5$ seconds. }
    \label{fig:kissat}
\end{figure}

Figs.~\ref{fig:minisat}, \ref{fig:kissat} show the TTS histogram of these two algorithms. Numerical fittings indicate that their TTS are heavy-tailed distributions whose tails are not exponentially bounded. Empirically, we can fit their TTS to a Weibull distribution,
\begin{equation}
    f(x)=
    \begin{cases}
        \frac{k}{\lambda}\left(\frac{x}{\lambda}\right)^{k-1}e^{-(x/\lambda)^k}, & x\geq 0, \\
        0, & x<0.
    \end{cases}\label{eq:weibull}
\end{equation}

\begin{figure}[htbp]
    \includegraphics[width=0.47\textwidth]{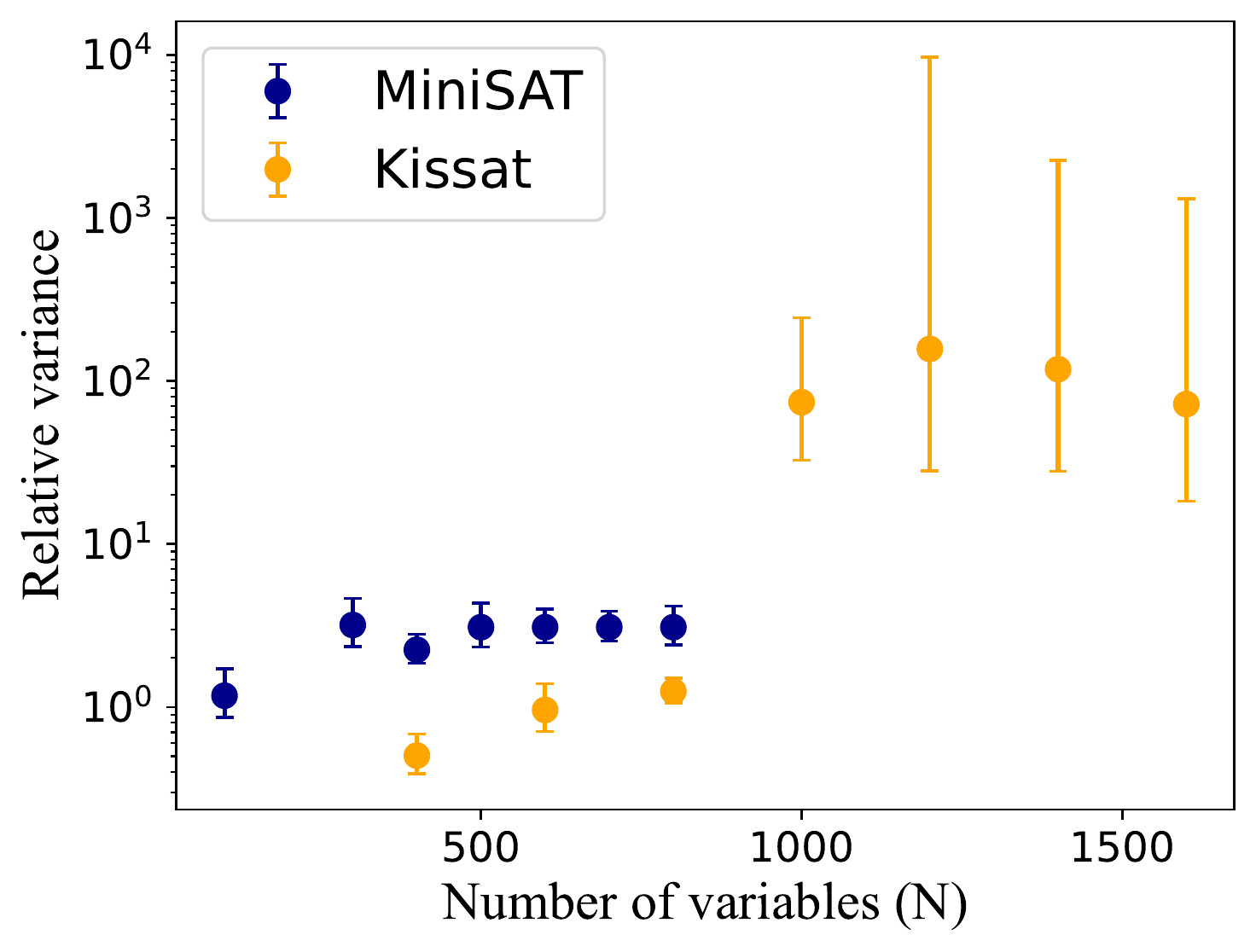}
    \caption{The relative variance, $\sigma^2(N)/\mu^2(N)$, of MiniSAT and Kissat. For both algorithms, as $N$ increases, the relative variance quickly grows first, then reaches a plateau. This confirms that these algorithms are also not self-averaging. }
    \label{fig:variance_CDCL}
\end{figure}

In Fig.~\ref{fig:variance_CDCL}, we plot the relative variance, $\sigma^2(N)/\mu^2(N)$, of both algorithms. Since not all instances are solved, the relative variances are computed using the fitted distributions. For both algorithms, the relative variance increases first, and then saturates when the number of variables is large enough. In other words, these algorithms are also not self-averaging. 

Note how the relative variance of the Kissat algorithm grows to extreme values. Although Kissat can solve most of the test instances in a short time, there are certain instances that take an extremely long time to solve. We suppose this phenomenon could be partially related to the goal of the SAT competition, which requires the solver to solve as many problems as possible within a limit of 5000 seconds. Loosely speaking, the ``no free lunch theorem'' \cite{wolpert1997no} states that, without any exploited structure, the computational cost of finding a solution averaged over all problems in the class is the same for any solution method. If a ``good'' solver can solve many instances within a very short time, this must mean that a certain portion of the instances will take an extremely long time to solve. Therefore, if a competition solver needs to solve more instances within a shorter time, a path is to ``amplify'' this difference across different instances, resulting in extremely large variances.

\section*{Self-averaging under different clause-to-variable ratios}

In the main text we used Barthel instances \cite{barthel2002} with clause-to-variable ratio 7. This choice is arbitrary, and different families of 3-SAT problems should lead to similar results. In this section, we present additional numerical results at ratio 6 to confirm this. 

\begin{figure}[htbp]
    \includegraphics[width=0.47\textwidth]{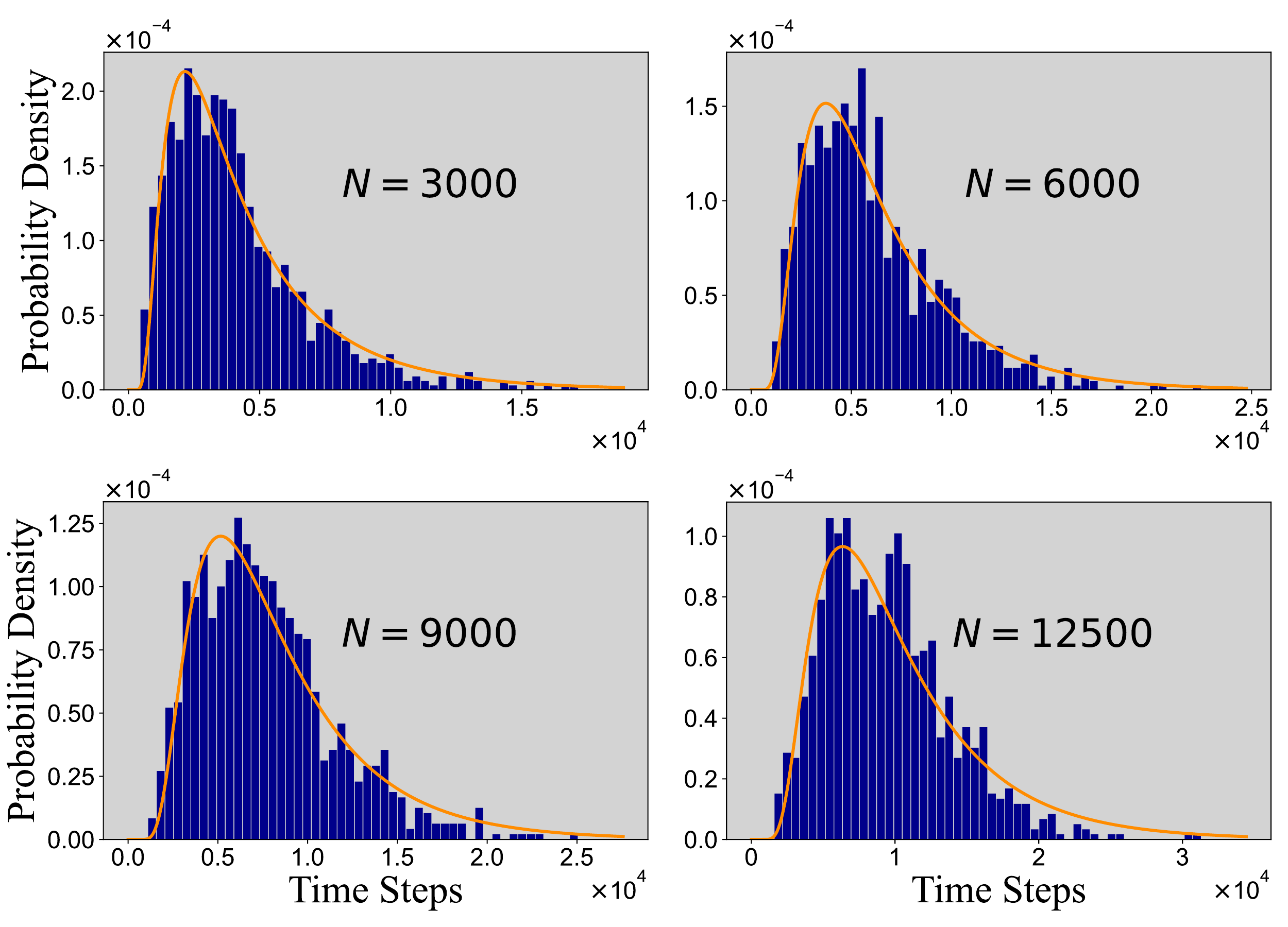}
    \caption{The distribution of TTS, solved using DMMs for Barthel instances with clause-to-variable ratio 6. Each histogram contains attempts on 1000 different instances, and all instances are successfully solved. Numerical integration is carried out using the forward Euler method, with a fixed time step $\Delta t=0.05$. }
    \label{fig:histogram_ratio6}
\end{figure}

In Fig.~\ref{fig:histogram_ratio6}, we show the histograms of TTS with varying number of variables $N$. In each histogram, 1000 problem instances are solved using DMMs, integrated with a fixed time step $\Delta t=0.05$ and noise strength $\Gamma=0.12$. All solution attempts are successful, and their TTS can be well-fitted using an inverse Gaussian distribution. 

\begin{figure}[htbp]
    \includegraphics[width=0.47\textwidth]{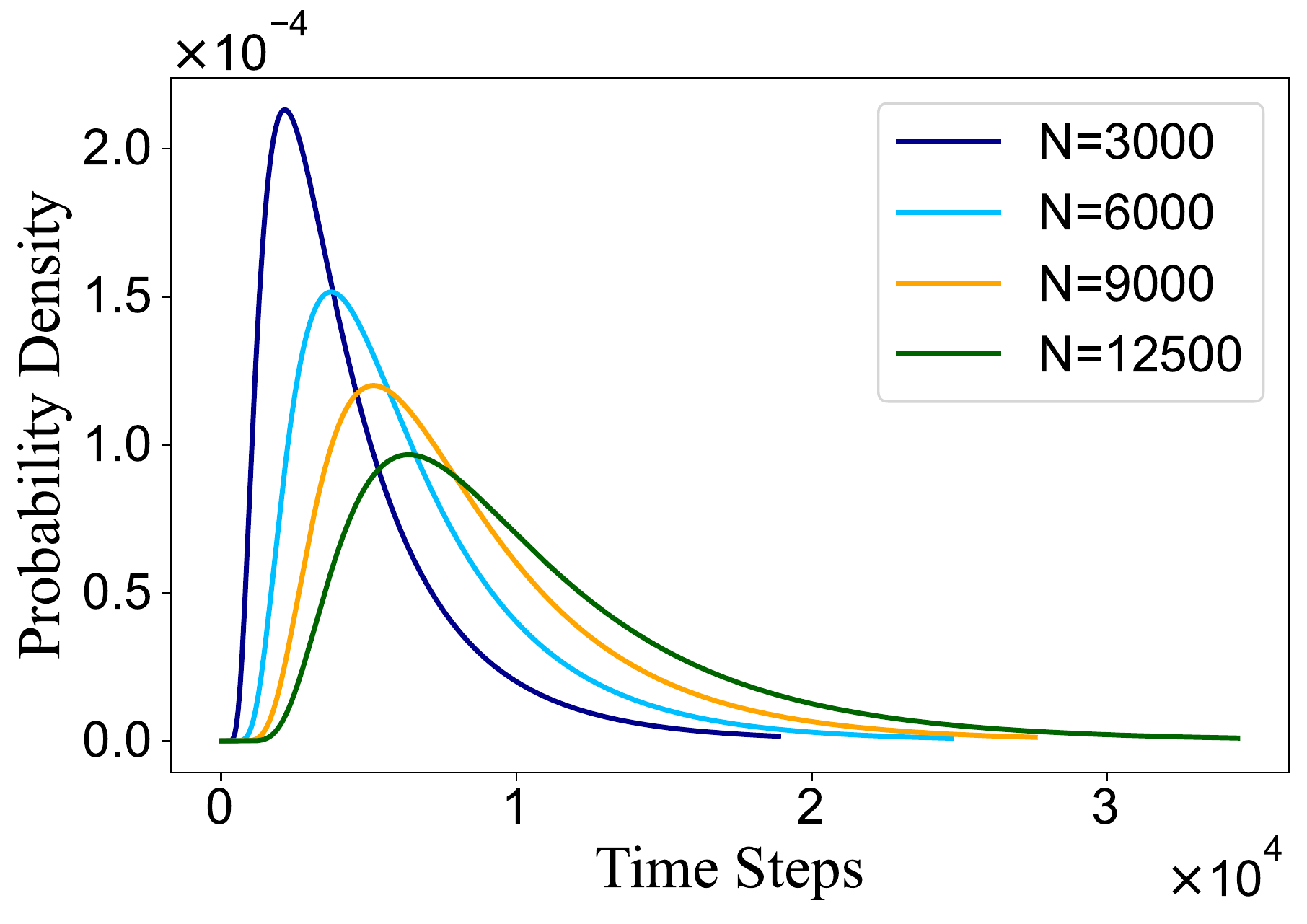}
    \caption{A comparison of the different fitted inverse Gaussian distributions with varying number of variables, $N$, at clause-to-variable ratio 6. The fits are taken from the histograms presented in Fig.~\ref{fig:histogram_ratio6}. }
    \label{fig:fits_ratio6}
\end{figure}

\begin{figure}[htbp]
    \includegraphics[width=0.47\textwidth]{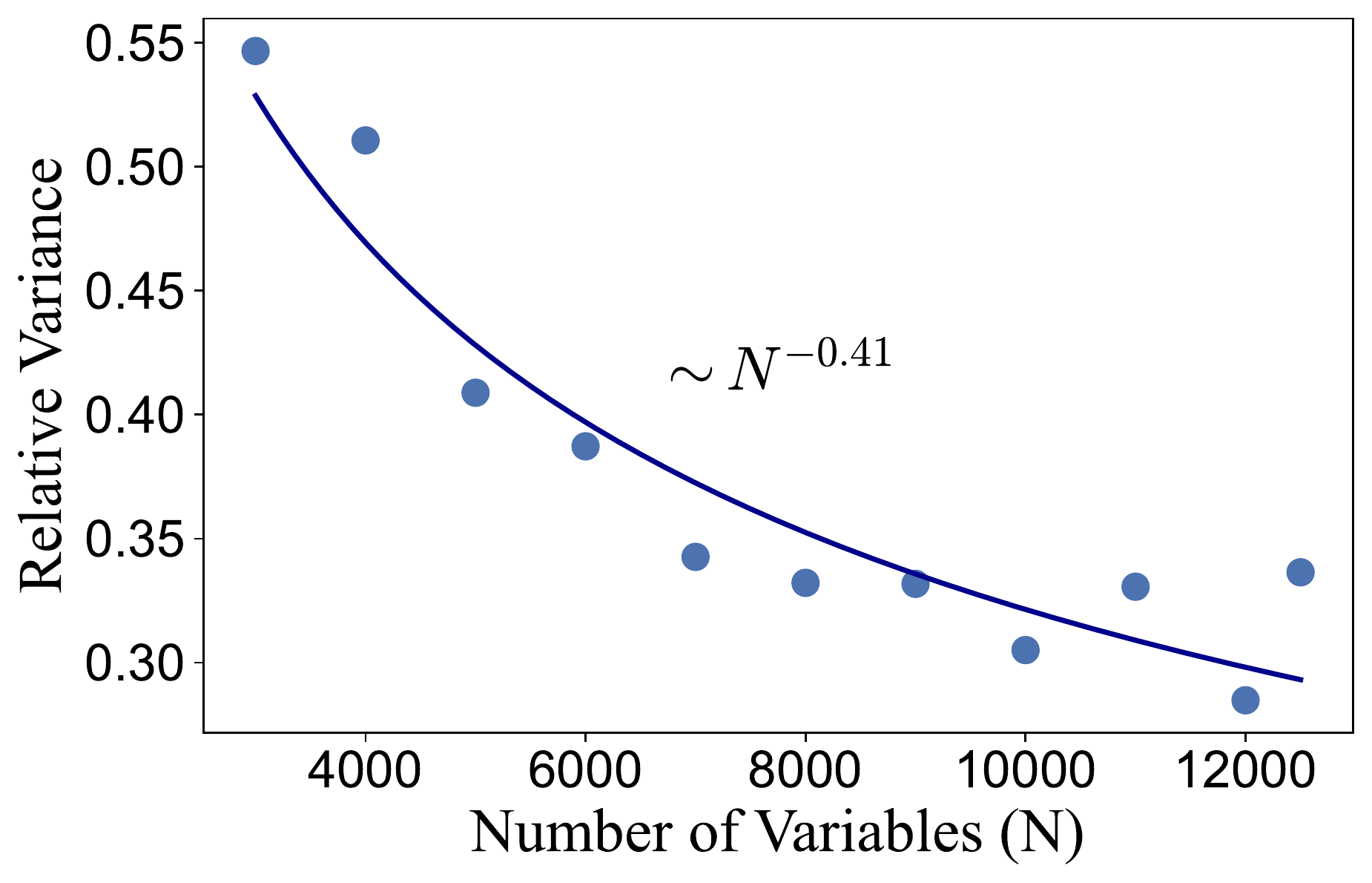}
    \caption{The relative variances, $\sigma^2/\mu^2$, as a function of the number of variables, $N$, at clause-to-variable ratio 6. The relative variance is well fitted with the power law $\sigma^2(N)/\mu^2(N)\sim N^{-0.41\pm 0.05}$, showing a weak self-averaging behavior. }
    \label{fig:variance_ratio6}
\end{figure}

A comparison of different TTS distributions with varying $N$ is shown in Fig.~\ref{fig:fits_ratio6}, and the relative variance, $\sigma^2(N)/\mu^2(N)$, is plotted as a function of $N$ in Fig.~\ref{fig:variance_ratio6}. A power-law fit indicates that $\sigma^2(N)/\mu^2(N)\sim N^{-0.41\pm 0.05}$, which is a weak self-averaging behavior, demonstrating that in this case numerical noise strongly couples with the additive noise. 

\section*{Self-averaging of TTS with only numerical noise}
\begin{figure}[htbp]
    \includegraphics[width=0.47\textwidth]{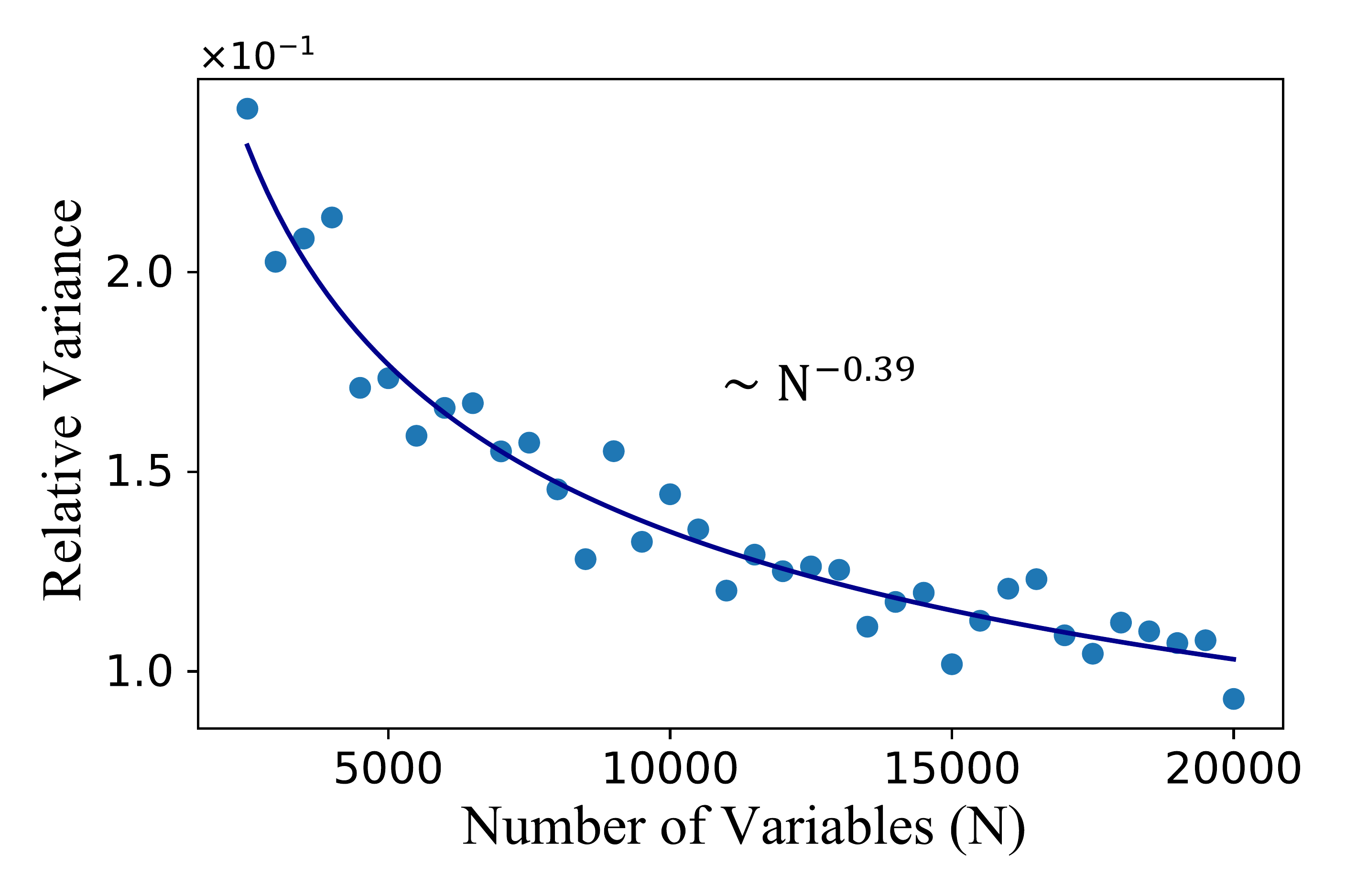}
    \caption{Self-averaging of the TTS for the DMM represented by the Eqs.~(2) and (3) in the main text integrated using forward Euler. For each size we have solved 1,000 3-SAT instances (from~\cite{barthel2002}) and the integration step used is $\Delta t= 0.2$ (in arbitrary units). The relative variance $\sigma^2(N)/\mu^2(N) \sim N^{-0.39\pm0.02}$, indicating a weak self-averaging. }
    \label{fig:variance_num}
\end{figure}

The numerical integration of Eqs.~(2) and (3) in the main text introduces errors which are akin to some sort of ``noise''. In fact, this type of noise is non-local, as it accumulates during the integration, while physical noise 
is typically local in both time and space. In addition, since the numerical noise enters the equations of motion in a non-linear way, it behaves differently than the additive physical noise we have considered in the main text. Nonetheless, in the presence of only numerical noise, we still observe both the inverse Gaussian probability distribution of the TTS as well as its self-averaging behavior. However, self-averaging is weaker than in the presence 
of physical additive noise and depends on the integration scheme used. In Fig. \ref{fig:variance_num} we show the relative variance over a range of 3-SAT instances ranging from 2,500 to 20,000 variables (at a fixed clause-to-variable ratio of 7) using the forward Euler integration scheme. This yields a relative variance of $\sigma^2(N)/\mu^2(N) \sim N^{-0.39\pm0.02}$, hence a weak self-averaging. 

\begin{figure}[htbp]
 	\includegraphics[width=0.47\textwidth]{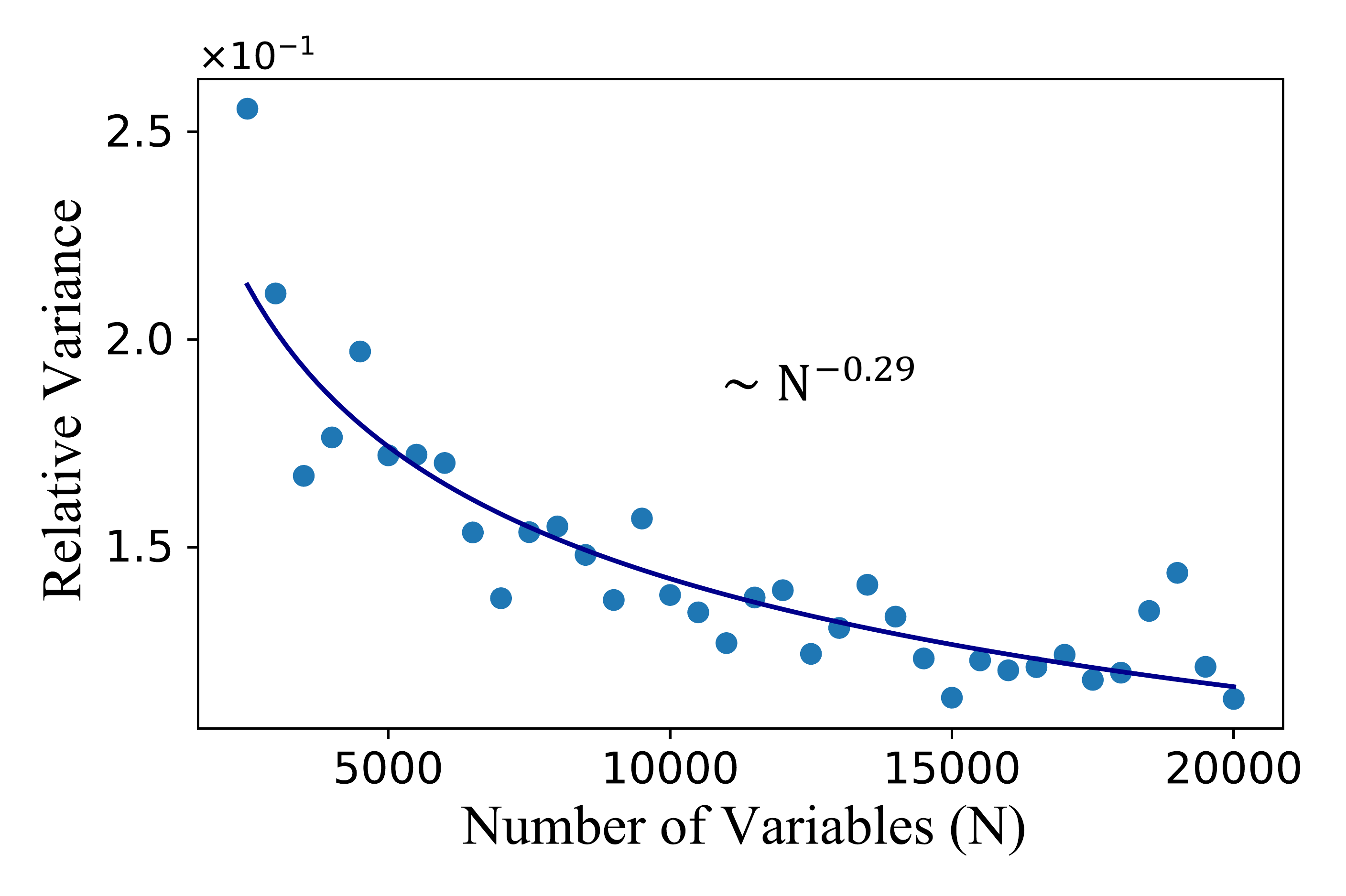}
 	\caption{Self-averaging of the TTS for the DMM represented by the Eqs.~(2) and (3) in the main text integrated using Runge-Kutta 4th order. For each size we have solved 1,000 3-SAT instances (from~\cite{barthel2002}) and the integration step used is $\Delta t= 0.2$ (in arbitrary units). The relative variance $\sigma^2(N)/\mu^2(N) \sim N^{-0.29\pm0.02}$, indicating a weak self-averaging different from the one obtained with the forward Euler integration scheme. }
 	\label{fig:variance_numRK4}
\end{figure} 
 
In Fig. \ref{fig:variance_numRK4}, we show the same calculation but with the Runge-Kutta 4th order integration scheme. 
The TTS is still weak self-averaging but with a different relative variance of $\sigma^2(N)/\mu^2(N) \sim N^{-0.29\pm0.02}$. This confirms our conclusions in the main text that in the presence of only numerical noise, the relative variance of the TTS goes to zero as $N^{-\theta}$ in the $N\rightarrow \infty$ limit, 
with $\theta <1$ and dependent on the integration scheme. 
 
\section*{Distribution of TTS on the same instance with random initial conditions}

\begin{figure}[htbp]
	\centering
	\includegraphics[width = 0.47\textwidth]{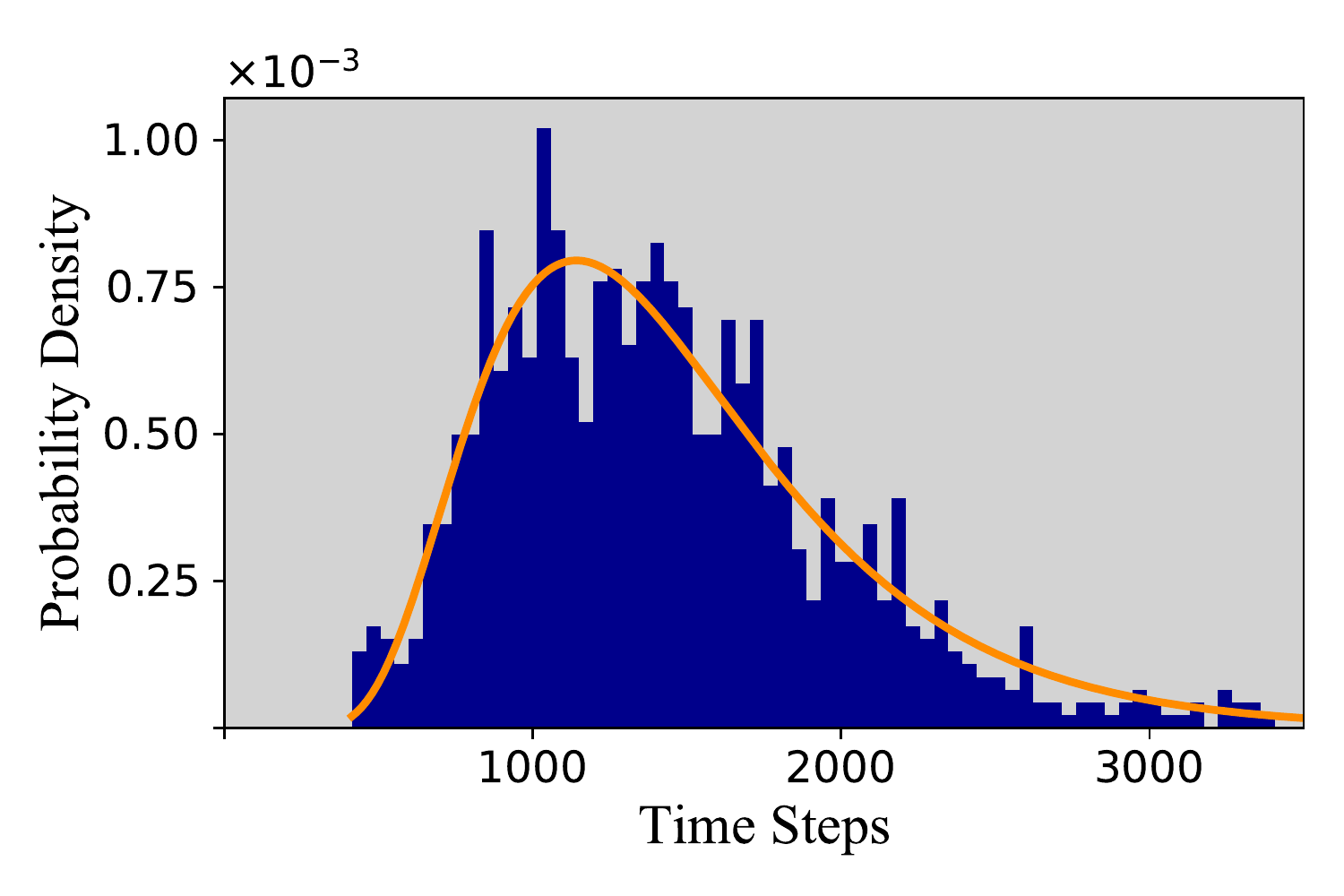}
	\caption{Histogram of TTS (in number of integration steps) of a DMM solving one 3-SAT instance taken from~\cite{barthel2002} with  $N =$ 6,000 variables and 42,000 clauses, with different, random initial conditions in each attempt. The histogram was generated by solving 981 out of 1,000 3-SAT attempts within a fixed time out. The fit is an inverse Gaussian distribution, Eq.~(5) in the main text, with 
		$\lambda = 6160\pm130$ and $\mu = 644\pm3$.}
	\label{fig:example_fit_same_instance}
\end{figure}

The simulations in the main text have been performed with different instances for each problem size and random initial condition per instance. In Fig. \ref{fig:example_fit_same_instance}, we show that solving the {\it same instance} many times with random initial conditions results in the same inverse Gaussian distribution we observe when different instances are solved in each attempt. This suggests that ``microscopic'' features of different 3-SAT problem instances appear to DMMs simply as different initial conditions in the same phase space. 

\end{document}